\documentclass[%
 reprint,
nofootinbib,
 amsmath,amssymb,
 aps,
]{revtex4-1}

\usepackage{graphicx,color}
\usepackage{dcolumn}
\usepackage{bm}

\usepackage{ulem}

\definecolor{cadmiumgreen}{rgb}{0.0, 0.42, 0.24}
\definecolor{blue(ryb)}{rgb}{0.01, 0.28, 1.0}

\begin{document}


\title{Exact critical exponents for vector operators in the 3d Ising model and conformal invariance}

\author{Gonzalo De Polsi}
\affiliation{Instituto de F\'isica, Facultad de Ciencias, Universidad de la
Rep\'ublica, Igu\'a 4225, 11400, Montevideo, Uruguay
}%

\author{Matthieu Tissier}%
\affiliation{Sorbonne Universit\'e, CNRS, Laboratoire de Physique Th\'eorique de
la Mati\`ere Condens\'ee, LPTMC, F-75005 Paris, France}%

\author{Nicol\'as Wschebor}
\affiliation{Instituto de F\'isica, Facultad de Ingenier\'ia, Universidad de la
Rep\'ublica, J.H.y Reissig 565, 11000 Montevideo, Uruguay
}%

\date{\today}

\begin{abstract}
It is widely expected that the realization of scale invariance in the critical regime implies conformal invariance for a large class of systems. This is known to be true if there exist no integrated operator which transforms like a vector under rotations and which has scaling dimension $-1$. In this article we give exact expressions for the critical exponents of some of these vector operators. In particular, we show that one operator has scaling dimension exactly $3$ in any space dimension. This operator turns out be the leading operator (i.e. the operator with the smallest scaling dimension) in $d=4$.  The nonrenormalization of this scaling dimension results from the fact that the associated operator is redundant. We explain why, contrarily to the common wisdom, it is important to deal with such operators in the present context. 
\end{abstract}

\maketitle


\section{\label{sec_intro}Introduction}

The vicinity of a second order transition is remarkable because the long-distance properties are invariant under dilatations, even though the underlying microscopic model involves some typical scales, such as a lattice spacing or a typical inter-particle distance. This emergent symmetry under dilatation is best described in the framework of the renormalization group. To each microscopic model one can associate an effective action $S_k$ which describes the dynamics of the coarse-grained model, where the short-distance degrees of freedom (as compared to the length scale $k^{-1}$) have been integrated out. Scale invariance shows up in this framework as a fixed point of the renormalization-group flow.

Soon in the '70s it was conjectured that other emergent symmetries may occur in the critical regime. In fact, it may be that the whole conformal group is realized \cite{Polyakov:1970xd,Migdal:1972tk}. This was proven to be true in bidimensional systems under quite general conditions \cite{Zamolodchikov:1986gt} but the situation is more intricate in the case of $d>2$.  The issue of the validity of conformal invariance
above two dimensions became of the utmost importance in the last few years given the success of the conformal bootstrap program in the Ising universality class \cite{ElShowk:2012ht,El-Showk:2014dwa,Kos:2014bka,Kos:2015mba}.

In Ref \cite{polchinski}, Polchinski showed that a model with translation, rotation and scale invariance also presents conformal invariance if there exists no virial current. A similar sufficient condition was derived in a different setting \cite{delamotte}, with slightly different prerequisites. In particular, Polchinski's sufficient condition requires the existence of a local energy-momentum tensor which is not always granted, as for example, in the long-range Ising model. The sufficient condition proposed in \cite{delamotte}
does not require the existence of a local energy-momentum tensor and therefore generalizes, for example, to the case of  mild long-range interactions.

In a nutshell, the derivation of the sufficient condition goes as follows \cite{delamotte}. We can write the Ward identity associated with conformal invariance in such a way that if conformal invariance is present, the right-hand side of the Ward identity vanishes. If conformal invariance is not realised, the right-hand-side, that we call $\Sigma_\mu$ is nontrivial. Now, it can be proven that $\Sigma_\mu$ is an eigenoperator of the renormalization-group flow, in the sense that 
$$\partial_t( \Gamma^\star+\epsilon_\mu \Sigma_\mu)\propto
\epsilon_\mu \Sigma_\mu+\mathcal O(\epsilon_\mu\epsilon_\nu)$$ where
$t$ is the renormalization-group time and $\Gamma^\star$ represents
the fixed point of the renormalization group.  Moreover, $\Sigma_\mu$
fulfills the following properties:
\begin{itemize}
    \item it transforms as a vector under space rotations,
    \item it is invariant under translations,
    \item it is a scalar under the internal symmetries of the problem (e.g. symmetric under $\phi\to-\phi$ in the Ising universality class),
   \item it has scaling dimension $-1$.
\end{itemize}

These sufficient conditions indicate a path to prove that conformal invariance is indeed realized in the critical domain of a particular system. We need to find a bound on the scaling dimensions of the vector operators with the properties described above. One strategy followed in \cite{delamotte} for the Ising universality class, on which we concentrate from now on, consists in using Griffiths and Lebowitz inequalities on correlation functions \cite{Griffiths, KellySherman, lebowitz74} in order to prove that any integrated vector operator invariant under $\mathbb{Z}_2$ symmetry has scaling dimension greater than $-1$. 

Another strategy consists in computing explicitly the lowest scaling dimension of such vector operators.
Several attempts have been performed in this direction during the last few years. In \cite{rychkov}, a Monte-Carlo simulation was performed with the aim of determining the scaling dimension of the vector operator $\int d^3x \phi\partial_\mu\phi(\partial_\nu\phi)^2$ appropriately discretized on a $3d$ lattice. The result quoted for the scaling dimension of the integrated vector operator is $3\pm 1$. Since the discretized operator considered in \cite{rychkov} is quite generic, it is natural to believe that it couples to the operator of lowest scaling dimension. Under this assumption, and invoking the aforementioned sufficient condition, this result strongly indicates that conformal invariance is present in the critical regime of the $d=3$ Ising model. From the analytic side, a 1-loop calculation \cite{delamotte} performed in $d=4-\epsilon$ shows that the integrated vector operator of lowest scaling dimension is $O_4^3=\int d^dx\, \phi^3\partial_\mu \Delta\phi$ where $\Delta$ is the Laplacian. This operator turns out to be the same (up to integration by parts) as
the one employed in the 3d case in Ref.~\cite{rychkov}. It has dimension $3+\mathcal O(\epsilon^2)$ (i.e., the correction linear in $\epsilon$ vanishes). These two results indicate that the smallest scaling dimension of an integrated vector operator is close to 3 in all dimensions. 

In Sect.~\ref{sec_nonrenorm}, we show that the scaling dimension
of the integrated vector operator studied previously is actually exactly 3 in any dimension. This eliminates the uncertainties coming from the numerical simulation in \cite{rychkov} and, accordingly, under the same assumptions made in that reference that scale invariance implies conformal invariance. 

As explained below, we can track back the fact that the scaling dimension of this operator can be computed exactly to the fact that it is a redundant operator. Such operators, first discussed by Wegner in \cite{Wegner}, are often discarded for several reasons. First, they lead to correlation functions which are short-ranged, which, at first sight, make them of little physical interest {\it per se}. For such operators, it is not possible to define a scaling dimension by studying the long-distance behavior of the correlation functions. We stress that it is nonetheless possible to define a scaling dimension by a stability analysis of the renormalization-group flow around the fixed point. Second, Wegner argued that the scaling dimensions of such operators actually change when nonlinear renormalization-group transformations are used. We show nevertheless that, in the context of testing whether conformal invariance is present in the critical regime, it is actually fundamental to deal with such redundant operators. We argue in Sect~\ref{sec_redundant} that proving that there exists no vector operator with the properties listed in the introduction, up to possible redundant operators is not sufficient for studying whether conformal invariance is valid or not. Instead, we show that it is fundamental to have control on the redundant operators too.

\section{Nonrenormalization theorem}
\label{sec_nonrenorm}
We consider a description of the Ising universality class in terms of continuous fields. We can choose the Hamiltonian (or action) to be of the
Ginzburg-Landau type:
\begin{equation}
\label{eq_action_micro}
	S[\phi]=\int_x \,\left[\frac{1}{2}(\nabla \phi)^2+\frac{1}{2}r_\Lambda\phi^2+\frac{u_\Lambda}
{4!}\phi^4\right],
\end{equation}
where $\int_x=\int d^d x$. We consider the model with an appropriate ultraviolet regulator at some scale $\Lambda$. In Eq.~(\ref{eq_action_micro}), the subscript $\Lambda$ indicates that the coupling constants are defined at the microscopic scale $\Lambda$. Following Polchinski and Wetterich \cite{Polchinski:1983gv,wetterich}, we add a quadratic regulator to the theory:
 \begin{equation}
\Delta S_{k}[\phi]=\frac{1}{2}\int_{x,y}\phi(x)R_k(|x-y|)\phi(y) \end{equation}
which regularizes the theory in the infrared. 
The properties of the regulating function are more conveniently discussed in Fourier space. The so-called regulating function $R_k(q)$  is chosen to approach zero exponentially fast  for $q\gg k$ and to saturate at a value which scales as $k^{2-\eta}$ when $q\ll k$. This ensures that the fluctuations of the long-distance modes (i.e. whose typical length scale are greater than $k^{-1}$) are effectively suppressed while the short-distance ones are kept unchanged. In what concerns the ultraviolet, we can regularize
the theory either by modifying the regulating function $R_k(q)$ \cite{Berges:2000ew} or by considering the model on a hypercubic lattice with lattice spacing $\pi/\Lambda$ at the price of introducing a discretization of the field derivatives. 

Following Wilson, a convenient strategy for determining the scaling dimension of an operator consists in studying the evolution of the corresponding coupling under the renormalization-group flow in the vicinity of the fixed point. To this end, we add to the action a part which couples to a vector operator:
\begin{equation}
\label{eq_actionV}
	S_{\text V}[\phi]=\int_x \frac{a^\mu_\Lambda}{3!}\phi^3\partial_\mu\Delta\phi=\int_x \frac{a^\mu_\Lambda}{4}(\partial_\nu\phi)^2\partial_\mu(\phi^2).
\end{equation}
where the last equality is obtained by integration by parts. The second expression corresponds to the form considered in \cite{rychkov}. \footnote{We stress that our proof relies only on {\it integrated} operators so that the two expressions are equally valid.} Moreover, it has been proved to be the most relevant integrated vector operator invariant under $\mathbb{Z}_2$ symmetry near $d=4$ \cite{delamotte}. Baring coincidences (or superselection rules) we expect this operator to couple to all $\mathbb{Z}_2$ symmetric vector operators, in particular to the most relevant one.

The critical Ising model is invariant under (space) rotations from which we conclude that, at the Wilson-Fisher fixed point, the dimensionless, renormalized, counterpart of $a^\mu$ vanishes. Moreover, since we are only interested in the scaling dimension of the vector operator around the Wilson-Fisher fixed point, we concentrate on infinitesimally small $a^\mu_\Lambda$.

The regularized partition function in presence of a source $J(x)$ then reads:
\begin{equation}
e^{W_{k}[J,a^\mu_\Lambda]}=\int\mathcal{D}\phi e^{-S-S_{\text V}-\Delta S_k+\int_x J\phi}
\label{eq_path}
\end{equation}
We now perform an infinitesimal transformation of the integration variable: $\phi\to\phi-a^\mu_\Lambda/u_\Lambda \partial_\mu\Delta \phi$ in the path integral. It is readily found that the quadratic pieces in the action, including the regulating term $\Delta S_k$, are invariant under this transformation. The variation of the quartic part of the action is found to compensate exactly $S_{\text V}$. We thus find that
\begin{equation}
	W_{k}[J,a^\mu_\Lambda]=W_{k}[J+\frac{a^\mu_\Lambda}{u_\Lambda}\partial_\mu\Delta J,0]+\mathcal O(a^\mu_\Lambda a^\nu_\Lambda)
\end{equation}
 We now introduce the scale dependent effective action as the (modified) Legendre transform \cite{wetterich}:
 \begin{equation}
	\Gamma_{k}[\phi,a^\mu_\Lambda]=-W_{k}[J,a^\mu_\Lambda]+\int_x J\phi-\Delta S_k[\phi]
\end{equation}
and check easily that 
 \begin{equation}
	\Gamma_{k}[\phi,a^\mu_\Lambda]=\Gamma_k[\phi+a^\mu_\Lambda\partial_\mu \Delta\phi,0]+\mathcal O(a^\mu_\Lambda a^\nu_\Lambda
).
\end{equation}
This last equation states that the evolution of the effective action with an infinitesimal $a^\mu_\Lambda$ is related to the effective action at vanishing $a^\mu_\Lambda$, up to a modification of the field.

This property can be used in the following way. Defining the running coupling constants $u_k$ and $a^\mu_k$ as the prefactors of, respectively, $\int_x \frac{1}{4!}\phi^4$ and $\int_x \frac{1}{3!}\phi^3\partial_\mu\Delta\phi$ in $\Gamma_k$, we obtain that $a^\mu_k/u_k$ is constant along the flow. To obtain the scaling dimension of the vector operator, we introduce dimensionless, renormalized quantities (denoted with tilde) as
\begin{align}
\tilde x&=k x\\
  \tilde \phi(\tilde x)  &=k^{-(d-2)/2}Z_k^{1/2}\phi(x),
\end{align}
where $Z_k$ scales as $ k^{-\eta}$ at the Wilson-Fisher fixed point with $\eta$ the anomalous dimension. 
The renormalized coupling constants are thus: 
\begin{align}
\tilde u_k&=k^{d-4} Z_k^{-2} u_k\\
    \tilde a_k^\mu&=k^{d-1} Z_k^{-2}  a_k^\mu.
\end{align}
At the critical point, $\tilde u$ flows to a fixed point value $u_\star$. Consequently, when $k\to 0$,
\begin{equation}
    \tilde a_k^\mu\sim a_\Lambda^\mu\frac{u_\star}{u_\Lambda}k^3
\end{equation}
which shows that the scaling dimension of $a^\mu$ is exactly~3.

The proof given above relies strongly on the particular microscopic action given in Eq.~(\ref{eq_action_micro}). This gives interesting non-universal information on the flow of the coupling $a_k^\mu$, but confers a preeminent role to the peculiar form of the Hamiltonian. To overcome this issue, we now present an alternative proof of the same result. To this end, we first recall the exact Wetterich flow equation \cite{wetterich} for the effective average action, expressed in terms of dimensionless, renormalized, fields:
\begin{equation}
\begin{split}
    \partial_t \Gamma_k[\tilde \phi]=&\int_{\tilde x} \frac{\delta \Gamma_k}{\delta \tilde \phi(\tilde x)}\left(\tilde x^\rho\partial_{\tilde x^\rho}+d_\phi\right)\tilde \phi(\tilde x)\\&+\frac 12\int_{\tilde x\tilde y}\partial_t\tilde R(\tilde x-\tilde y)\tilde P_k(\tilde x,\tilde y)
\label{eq_flow}
\end{split}
\end{equation}
\vspace{.5cm}
where $ R_k( x)=Z_k k^{d+2}\tilde R(k x)$, $d_\phi=(d-2+\eta)/2$, $t=\log(k/\Lambda)$ and $\tilde P_k$ is the dimensionless, renormalized, propagator:
\begin{equation}
\int_{\tilde y}    \tilde P_k(\tilde x,\tilde y)\left[\frac{\delta^2\Gamma_k}{\delta\tilde \phi(\tilde y)\delta\tilde \phi(\tilde z)}+\tilde R(\tilde y-\tilde z)\right]=\delta(\tilde x-\tilde z)
\end{equation}

We now identify an exact eigenvector of the linearized flow. To this end, we add to the Wilson-Fisher fixed-point effective action $\Gamma_\star$ a small perturbation
\begin{equation}
    \label{eq_pert}
    \Gamma_k=\Gamma_\star+\tilde r_\mu(t)\int_{ \tilde x}
    \frac{\delta\Gamma_\star}{\delta \tilde \phi(\tilde x)} \tilde
    \partial_\mu\tilde\Delta \tilde \phi(\tilde x)
\end{equation}
 and we compute the flow of this functional at linear order in $\tilde r_\mu$.
\begin{widetext}
\begin{equation}
    \begin{split}
\partial_t{\tilde r}_\mu(t) \int_{\tilde x}\frac{\delta\Gamma_\star}{\delta \tilde \phi(\tilde x)} \tilde{\partial}_\mu\tilde{\Delta} \tilde \phi(\tilde x)&=
\tilde r_\mu(t)\int_{\tilde x\tilde y} \frac{\delta \ }{\delta \tilde \phi(\tilde y)}\left[\frac{\delta\Gamma_\star}{\delta \tilde \phi(\tilde x)} \tilde\partial_\mu\tilde\Delta \tilde \phi(\tilde x)\right]\left(\tilde y^\rho\partial_{\tilde y^\rho}+d_\phi\right) \tilde \phi(\tilde y)\\&-\frac12 \tilde r_\mu(t)\int_{\tilde x\tilde y\tilde z\tilde v\tilde w}\partial_t\tilde R(\tilde x-\tilde y)\tilde P_\star(\tilde y,\tilde z)\Gamma_\star^{(3)}(\tilde z,\tilde v,\tilde w)\tilde P_\star(\tilde w,\tilde x)\tilde\partial_\mu\tilde\Delta\tilde \phi(\tilde v)
    \end{split}
\end{equation}
On the other hand, if we derive the fixed point equation with respect to $\tilde \phi(\tilde x)$, multiply by $ \tilde\partial_\mu\tilde\Delta \tilde \phi(\tilde x)$ and integrate over $\tilde x$, we get
\begin{equation}
    \begin{split}
        0&=\int_{\tilde x\tilde y}\tilde\partial_\mu\tilde\Delta \tilde \phi(\tilde x)\frac{\delta \ }{\delta \tilde \phi(\tilde x)}\left[\frac{\delta\Gamma_\star}{\delta \tilde \phi(\tilde y)} \left(\tilde y^\rho\partial_{\tilde y^\rho}+d_\phi\right)\tilde \phi(\tilde y)\right]-\frac12 \int_{\tilde x\tilde y\tilde z\tilde v\tilde w}\partial_t\tilde R(\tilde x-\tilde y)\tilde P_\star(\tilde y,\tilde z)\Gamma_\star^{(3)}(\tilde z,\tilde v,\tilde w)\tilde P_\star(\tilde w,\tilde x)\tilde\partial_\mu\tilde\Delta\tilde \phi(\tilde v)
    \end{split}
\end{equation}
\end{widetext}
Combining the two equations, we obtain:
\begin{equation}
    \begin{split}
\partial_t {\tilde r}_\mu(t) \int_{\tilde x} \frac{\delta\Gamma_\star}{\delta \tilde\phi(\tilde x)} &\tilde\partial_\mu\tilde\Delta \tilde\phi(\tilde x)=\\
&\tilde r_\mu(t)\int_{\tilde x} \frac{\delta\Gamma_\star}{\delta \tilde \phi(\tilde x)} \left[\tilde \partial_\mu\tilde \Delta ,\tilde x^\rho\partial_{\tilde x^\rho}\right] \tilde\phi(\tilde x)
    \end{split}
    \label{proof_EV}
\end{equation}
The commutator is easily evaluated to be equal to $3\tilde \partial_\mu\tilde \Delta $. From this we deduce that the small perturbation introduced in Eq.~(\ref{eq_pert}) is an exact eigenoperator of the flow around the fixed point, with eigenvalue 3. This is consistent with the result found in the one-loop calculation of \cite{delamotte}, that we reproduce in the appendix \ref{app_1loop} for completeness. It is also consistent with the Monte-Carlo simulation performed in $d=3$ \cite{rychkov}.

We can generalize the previous result in different ways. First, we can change the power of the Laplacian in Eq.~(\ref{eq_pert}) from unity to a positive integer $n$. The main change appears at the level of Eq.~(\ref{proof_EV}), where the commutator is now $[\tilde\partial_\mu\tilde\Delta^n ,\tilde x^\rho\partial_{\tilde x^\rho}]=(2n+1)\tilde\partial_\mu\tilde\Delta^n$. This implies that the associated eigenvector has dimension $2n+1$. As a check of this result, we have considered the vector eigenoperators compatible with the $\mathbb{Z}_2$ symmetry whose scaling dimensions are 5 in $d=4$ and we have computed their first correction in $\epsilon=4-d$. There are four (independent) such operators: one ($O_6^3$) with 6 powers of the field and 3 derivatives and three ($O_{4,i}^5$ with $i\in\{1,2,3\}$) with 4 powers of the field and 5 derivatives. A one-loop calculation shows that $O_6^3$ has scaling dimension $5-5\epsilon/3+\mathcal O(\epsilon^2)$. The eigenvectors $O_{4,i}^5$ have dimensions $5+\mathcal O(\epsilon^2)$, $5-4\epsilon/9+\mathcal O(\epsilon^2)$ and $5-2\epsilon/3+\mathcal O(\epsilon^2)$. The eigenoperator with scaling dimension $5+\mathcal O(\epsilon^2)$ is found to be $\int_{\tilde x} \tilde \phi^3 \tilde \partial_\mu\tilde \Delta^2\tilde \phi$, in agreement with the general result mentioned above. Other relations can be obtained if we consider in Eq. (\ref{eq_pert}) an odd number of derivatives, with Lorentz indices not necessarily contracted together.

The present result also generalizes to the long-range Ising model, where the interaction between spins is not limited to nearest neighbors but decay as a power-law:
\begin{equation}
    H=-\sum_{i,j}J(i-j)S_i S_j
\end{equation}
where $J(i-j)\sim |i-j|^{-d-\sigma}$ and $\sigma$ is the exponent characterizing the decrease of the interactions. When $0<\sigma<2-\eta$, the model still has an
extensive free-energy but belongs to a different universality class than the local Ising model. The Ginzburg-Landau Hamiltonian is identical to the one given in Eq.~(\ref{eq_action_micro}) except that the quadratic part is now, in Fourier space,
\begin{equation}
    \int \frac{d^dq}{(2\pi)^d}\phi(-q) q^{\sigma}\phi(q).
\end{equation}
It is easy to verify that all the present analysis still applies to this case. We have checked that the one-loop calculation around the upper critical dimension
$d_c=2\sigma$ gives that the most relevant integrated vector operator has scaling dimension $3+\mathcal{O}(\epsilon^2)$. This result is important because
it justifies the use and the surprising success of the conformal bootstrap program in this model \cite{Paulos:2015jfa}.

We can also generalize the result to other internal groups. For O(N) theories, an exact eigenoperator can be found by adding a common O($N$) index on both the functional derivative and the field appearing in Eq.~(\ref{eq_pert}) and summing over this index. The associated eigenvalue is again 3 (or $2n+1$, if we change the power of the Laplacian). In \cite{delamotte}, we computed the scaling dimensions of the two vector operators of lowest dimension in an expansion in $\epsilon$ and found $3+\mathcal{O}(\epsilon^2)$ and $3-6\epsilon/(N+8)+\mathcal{O}(\epsilon^2) $. This result is consistent with the nonrenormalization theorem proven here. Let us stress, however, that in the O(N) model the non renormalization theorem does not constraint the leading vector operator but the next-to-leading, as can be seen already at one-loop level \cite{delamotte}.

\section{Redundant operators}
\label{sec_redundant}
 The operator appearing in $S_V$, see Eq.~(\ref{eq_actionV}), is obtained by computing the infinitesimal variation of the action under a modification of the field $\phi\to \phi+\epsilon \delta \phi$. Operators obtained in this way are called redundant operators and are sometimes considered to be physically uninteresting. Indeed, they typically have short-range correlation functions.\footnote{This implies that it is not possible to define their scaling dimension by looking at the power-law behavior of correlation functions at long distances. It is however possible to define a {\it bona fide} scaling dimension by a stability analysis of the renormalization-group flow around the fixed point. We stress that the constraint given in the introduction on the scaling dimension for the operator $\Sigma_\mu$ corresponds to this latter definition.} Moreover, Wegner showed that their scaling dimension can change when nonlinear scheme transformations are performed in the renormalization-group equation. 

The aim of this section is to show that, despite these facts, we {\it need} to consider these redundant operators if we want to prove that conformal invariance is realized in the critical domain. We should first stress that we do not consider here vector operators because of their intrinsic physical interest, but as possible candidates for inducing a breaking of conformal invariance. In principle, a redundant operator, with short-range correlations could be responsible for the breaking of conformal invariance. The mere existence of such an operator would have strong physical consequences because correlation functions for other fields would not display conformal invariance. We illustrate this argument in a simpler situation. Consider a model with two scalar fields $\phi_1$ and $\phi_2$ whose dynamics is given by a general action $S$ which needs not be O(2)-symmetric. If we perform, in the path integral of the partition function, a change of variable $\phi_i\to\phi_i+\theta \epsilon_{ij}\phi_j$ (here $\epsilon_{ij}$ is the bidimensional Levi-Civita tensor and $\theta$ an infinitesimal angle) which corresponds to an infinitesimal rotation in internal space, we obtain:
\begin{equation}
\label{eq_o2}
    \int d^dx\left(\epsilon_{ij}J_i\frac{\delta W}{\delta J_j}\right)=\int d^dx\left\langle \epsilon_{ij}\phi_i \frac{\delta S}{\delta \phi_j}\right\rangle.
\end{equation}
The brackets in the right-hand-side represent an average over the fields with the Boltzmann distribution in presence of sources $J_i$ for the fields $\phi_i$. Of course, if the action is O(2) symmetric, we recover the Ward identity for rotation in internal space. However, for a generic action $S$, the right-hand-side does {\it not} vanish and the O(2) Ward Identity is not satisfied. Now, what is of interest for us here is that the right-hand-side of the previous equation is the average of a {\it redundant} operator. The operator $\epsilon_{ij}\phi_i \frac{\delta S}{\delta \phi_j}$ appearing in the right-hand-side of Eq.~(\ref{eq_o2}), which has only contact terms in its correlation functions, is physically important because it induces a breaking of O(2) invariance, at the level of Ward identities.

To make an analogy with the strategy followed in this article to study conformal invariance, suppose we want to prove that a model is invariant under O(2) by searching for putative operators that could appear in the right-hand-side of Eq.~(\ref{eq_o2}). Suppose that we can discard the existence of such operators which are not redundant but that we have no control on redundant ones. Then, the previous example shows that we have no way to conclude on the O(2) invariance of the theory. If, instead, we can discard both non-redundant and redundant operators, then we conclude that the theory is indeed invariant.\footnote{It is often stated in the literature that redundant operators can be reabsorbed by a change of variables and are therefore not physically relevant. This however cannot be applied as such when testing whether a Ward identity is valid or not. Indeed, this would lead us to the absurd conclusion that a generic theory with two scalar fields can always be made O(2)-invariant by reabsorbing the redundant operator appearing in the right-hand-side of Eq.~(\ref{eq_o2}) through a field redefinition.}

The situation is slightly more complex for conformal invariance because  conformal symmetry, as dilatation symmetry, is valid only at long distances. The microscopic action does not present this symmetry. Apart form this extra complication, the argument given above for O(2) invariance applies for conformal invariance. If we can discard the existence of vector operators $\Sigma_\mu$ with the properties listed in the introduction, except for possible redundant operators, we have no way to conclude on the fate of conformal invariance. In order to prove that conformal invariance is realized, we have to discard also the redundant operators. 

The second criticism one may object to redundant operators is that their scaling dimension may change when the renormalization-group equation is modified. In this respect, we should stress that our derivation of the scaling dimension of the operator $\Sigma_\mu$ relies on a particular renormalization-group equation and it might well be that the condition for the scaling dimension of $\Sigma_\mu$ changes also when we modify the renormalization-group equation. This question is out of the scope of this article and we leave it to further work.

\section{Conclusion}
To conclude, we have shown that there exists a family of eigenoperators which transform as vectors under rotations, are scalars under the internal group and whose scaling dimension receive no loop correction. Among these operators lies the operator previously analyzed in a Monte-Carlo simulation in $d=3$ \cite{rychkov}, $O_4^3=\int_x\phi^3\partial_\mu \Delta\phi$ which has the lowest scaling dimension in $d=4$. This result is interesting at the light of the sufficient condition under which scale invariance implies conformal invariance, mentioned in the introduction. Indeed, as long as we admit (as is usually assumed) that this operator couples to the leading integrated operator, its scaling dimension being larger than $-1$, we would have an alternative proof to the one of \cite{delamotte} that scale invariance implies conformal invariance in all dimensions for the Ising universality class.

\appendix
\section{One-loop calculation of the scaling dimension}
\label{app_1loop}
In this section, we describe the calculation of the scaling dimension of the dominant vector operator at one loop. Using as a microscopic action the sum of $S$ and $S_V$ given respectively in Eqs.~(\ref{eq_action_micro}) and (\ref{eq_actionV}), we get a 4-point vertex of the form:
$$S^{(4)}(p_1,p_2,p_3)=u_\Lambda+ia_\Lambda^\mu\sum_{i=1}^4p_i^\mu(p_i)^2$$
where $p_4=-p_1-p_2-p_3$ is fixed by momentum conservation. We then compute the divergent part of the 1PI 4-point vertex, at 1 loop. Our aim here is to compute the scaling dimension of the vector operator by a stability analysis around the Wilson-Fisher fixed point, characterized by $a_\Lambda^\mu=0$. It is therefore sufficient to keep the terms linear in $a_\Lambda^\mu$. The calculation proceeds as usual. We introduce counterterms for $u$ and $a$ and derive from these the beta functions for the associated renormalized operators. We get:
\begin{align}
\beta_u&=   - \epsilon u +\frac 3{16\pi^2} u^2+\mathcal{O}(u^3)\\
\beta _{a_\mu}&= \left[(3-\epsilon)+\frac{3}{16\pi^2} u+\mathcal{O}(u^2) \right]a_\mu+\mathcal{O}(a_\mu a_\nu)\label{betaa}
\end{align}
Replacing the running coupling constant by its fixed-point value in Eq.{(\ref{betaa})}, we find that the 1-loop correction to $\beta_{a_\mu}$ exactly compensates the dimensional contribution $-\epsilon$. This shows that, at 1-loop, the scaling dimension of the vector operator is not renormalized and takes value $3+\mathcal O(\epsilon^2)$.

\begin{center}{\bf Acknowledgments}
\end{center}
The authors acknowledge Bertrand Delamotte, Nicolas Dupuis, Tim Morris and Kay Wiese for useful discussions. This work received support from Grant No. 412 FQ 293 of the CSIC (UdelaR) commission and Programa de Desarrollo de las Ciencias B\'asicas (PEDECIBA), Uruguay and ECOS sud U17E01.

\end{document}